\documentclass[aps,pra,twocolumn,english,showpacs]{revtex4-1}
\usepackage{siunitx} 
\usepackage{graphicx}
\usepackage{epstopdf}
\graphicspath{ {./Images/} }
\usepackage{amssymb}
\usepackage{amsmath}
\usepackage{centernot}
\usepackage{siunitx} 
\usepackage[usenames, dvipsnames]{color}

\usepackage{hyperref}
\hypersetup{
  colorlinks   = true, 
  urlcolor     = blue, 
  linkcolor    = blue,
  citecolor    = blue 
}

\newcommand{\ket}[1]{|#1\rangle}

\begin{document}

\title{State-dependent fluorescence of neutral atoms in optical potentials}

\author{M. Martinez-Dorantes, W.Alt, J. Gallego, S. Ghosh, L. Ratschbacher, and D. Meschede}
\affiliation{Institut f\"ur Angewandte Physik der Universit\"at Bonn, Wegelerstrasse 8, 53115 Bonn, Germany}

\begin{abstract}  
Recently we have demonstrated scalable, non-destructive, and high-fidelity detection of the internal state of $^{87}$Rb neutral atoms in optical dipole traps using state-dependent fluorescence imaging [M. Martinez-Dorantes et al., PRL, 2017]. In this article  we provide experimental procedures and interpretations to overcome the detrimental effects of heating-induced trap losses and state leakage. We present models for the dynamics of optically trapped atoms during state-dependent fluorescence imaging and verify our results by comparing Monte Carlo simulations with experimental data. Our systematic study of dipole force fluctuations heating in  optical traps during near-resonant illumination shows that off-resonant light is preferable for state detection in tightly confining optical potentials.
\end{abstract}


\maketitle

\section{Introduction}

Spatially resolved detection of individual neutral atoms in optical potentials has paved the way towards observation of quantum many-body phenomena all the way down to the individual atom level. For many of the potential applications in the field of quantum simulation and quantum information processing with cold neutral atoms however, it is essential not only to record spatial distributions of atoms but also to detect their internal qubit states in situ. In current experiments the detection of the qubit state of multiple atoms, encoded in their hyperfine ground state, is achieved by removing atoms in one of the qubit states from the optical trap with a resonant light pulse; Subsequently the remaining atoms are imaged in a state-insensitive way~\cite{kuhr2003}. This so-called \emph{push out} method achieves high fidelities for arrays of neutral atoms, but has the disadvantage of being intrinsically destructive. The reloading of atoms lost from their optical trapping potentials, for example, limits the duty cycle of experiments and fully hinders the application of feedback in quantum error correction algorithms. \\ \indent
Efficient single-shot readout without atom losses has previously been achieved for single atoms coupled to high-finesse optical cavities~\cite{gehr2010,bochmann2010,reick2010,Khudaverdyan2009}. More recently this has been extended to individual atoms trapped by optical tweezers in free-space using single photon counters for state-selective fluorescence detection~\cite{fuhrmanek2011,gibbons2011} and in state-dependent optical lattices where the internal state are mapped onto the atom's position~\cite{2016RobensAtomicBomb}. \\ \indent
In this article we investigate the processes underlying state-selective fluorescence that has recently been used to demonstrate fast, simultaneous and non-destructive state imaging of neutral atoms~\cite{dorantes2017,kwon2017}. We show that, contrary to usual detection settings, off-resonant light is preferable over resonant light when the atoms are trapped in tightly confining optical potentials.\\ \indent

\begin{figure*}[t]
\centering
    \includegraphics[trim={9cm 0 8cm 0},width=0.5\columnwidth]{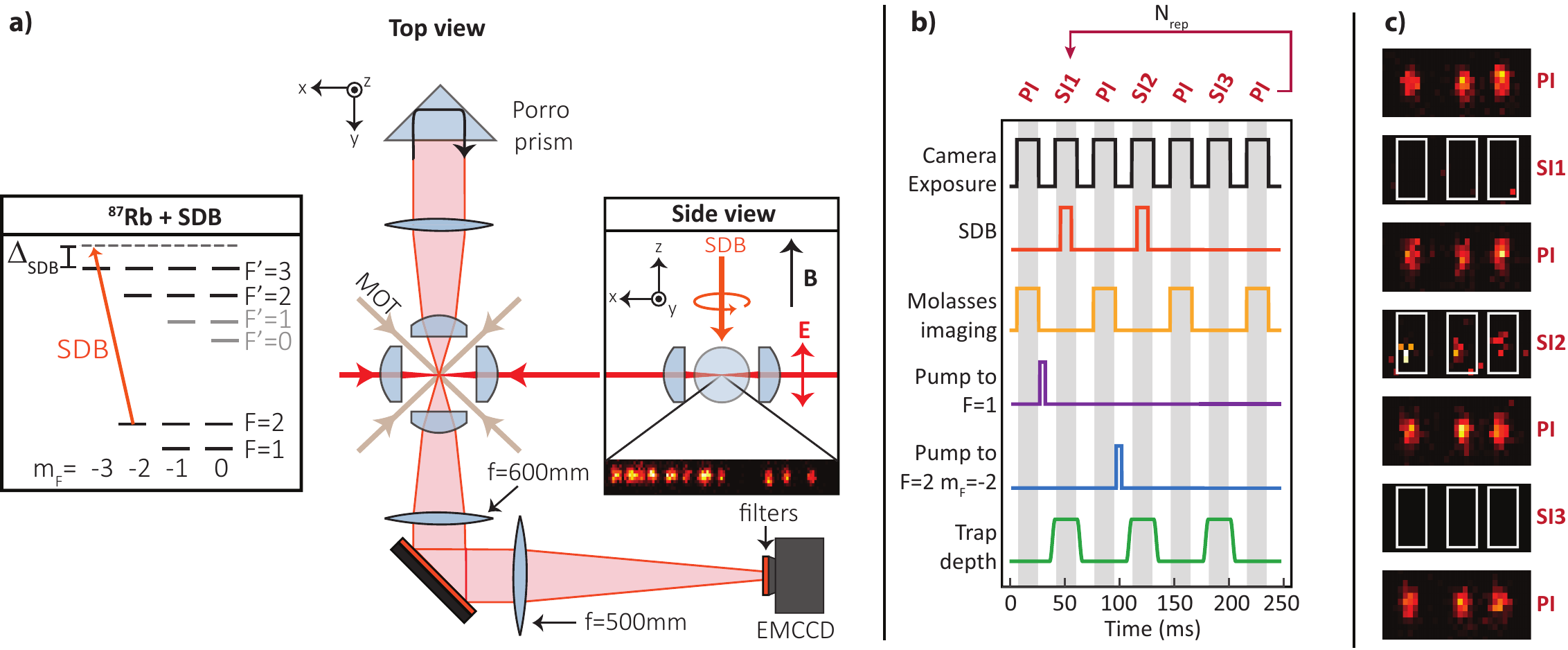}
  	\caption[Setup] {Experimental setup and measurement sequence. \textbf{a)}~Schematic view of the experimental setup with propagation directions and polarizations of the various light fields, and the orientation of the magnetic (B-) field. For state-dependent imaging only a single near-resonant \emph{ state detection beam} (SDB) is employed to scatter photons on the $\ket{2,-2} \rightarrow \ket{3',-3'}$  cycling transition. Fluorescence light from the atoms is collected by two opposing lenses and reflected by a Porro prism used as to form a single overlapped image on an EMCCD camera. \textbf{b)}~Diagram of the experimental sequence used to characterize the atom survival probability and the mean error of the state detection (see text for details). \textbf{c)}~ Typical set of images acquired by the EMCCD camera for \emph{position detection imaging}~(PI) and \emph{state detection imaging}~(SI) during one iteration of the experimental sequence. For the purpose of state detection, the counts in images acquired during SI are integrated for individual atoms over the respective region of interest (white boxes).} {\label{fig:fig1}}
\end{figure*}
\subsubsection{State detection by fluorescence imaging}
\label{introFluorescence}
State-selective fluorescence consists of addressing a cycling transition of an atom with laser light, such that it will scatter a considerable amount of photons, if the atom was initially in the \emph{bright} (B) qubit state, and practically no photons if it started in the \emph{dark} (D) qubit state. If a sufficient number of  scattered photons is detected, the original state of the atom can be inferred by statistical analysis. \\ \indent
In practice, the ideal case of well-separated counting distributions for bright and dark state atoms is impaired by several imperfections: leakage between the \emph{bright} and \emph{dark} state due to off-resonant light scattering fundamentally constrains the  state determination fidelity for a given photon detection efficiency and atomic level structure. In contrast to  trapped ions, where state detection fidelities beyond the requirements of fault-tolerant error correction are routinely achieved~\cite{acton2006,burrell2010}, neutral alkali atoms lack high-lying metastable states that allow for electron shelving techniques. 
For neutral atoms, commonly trapped in  shallow optical potentials, also the motional dynamics caused by near-resonant scattering of light plays a decisive role: The balance of heating and cooling processes caused by photon recoil, differences in the trap potentials of ground and excited states,  and Doppler and sub-Doppler cooling effects determine the time at which atoms are lost from the trap during the illumination process and thus how many photons can be detected. \\

\subsubsection{Structure of the article}

In this article we study the coupled dynamics of the internal and external degrees of freedom during photon scattering. In Sec.~\ref{experimentalsetup} we introduce the experimental setup for state-dependent fluorescence imaging of neutral atoms. In Sec.~\ref{HeatingDynamics} the model of atom cooling and heating dynamics in optical potentials induced by near-resonant light scattering  is presented.  Analytic and theoretical results obtained by Monte Carlo simulations are validated with measurements for a range of parameters. Sec.~\ref{StateDependentImaging} focuses on the details of  state detection based on fluorescence imaging, including population leakage and detector imperfections, and Sec.~\ref{Conclusions} concludes with a discussion about the applicability of our results to other cold atom experiments. In the Appendix, the measurement of the experimental photon detection efficiency, and the algorithm of the Monte Carlo simulation are described in more detail.

\section{Experimental Setup}
\label{experimentalsetup}

In the experiment, neutral $^{87}$Rb atoms are confined in a red-detuned optical dipole trap at a wavelength of $\lambda_\text{DT}= \SI{860}{\nano\meter}$. The optical potential with a maximal trap depth for the electronic ground states of $U_0=k_{\rm{B}} \times \SI{3.5}{\milli\kelvin}$ (with $k_{\rm{B}}$ the Boltzmann constant) is a standing-wave formed by two counter-propagating laser beams focused to a waist radius of $\SI{4.8}{\micro\meter}$. The atoms are directly transferred into the dipole trap from  a small magneto-optical trap (MOT) loaded from Rubidium background vapor in the ultra-high-vacuum chamber (see Fig.~\ref{fig:fig1}a). The initial spatial distribution of atoms in the trap is then compressed~\cite{depue1999unity} to within the Rayleigh length of the dipole trap beams to reduce the inhomogeneity of the potential depths experienced by the atoms.\\ \indent
Optical molasses illumination by the MOT cooling and repumping beams cools the atoms in the dipole trap to a steady-state temperature of about $\SI{80}{\micro\kelvin}$. Their fluorescence light is collected by two in-vacuum, high numerical aperture, aspheric lenses and imaged onto an EMCCD camera with a measured magnification of 35.4  and a total photon detection efficiency of $2.9\%$ (see Appendix A). The high signal-to-noise ratio image is used for precise determination of the atoms positions in the lattice (\emph{position detection imaging,}~PI)~\cite{Alberti2016SuperResolution}. \\ \indent

In order to realize \emph{state detection imaging}~(SI) that is sensitive to the initial hyperfine state, i.e. bright images for atoms initially in the $\ket{F\!=\!2, m_F\!=\!-2}$ state and dark images for atoms in the $F\!=\!1$ hyperfine manifold, the atoms are illuminated by  a single \emph{state detection beam} (SDB). The beam is is circularly polarized and tuned near the $\ket{2,-2} \rightarrow \ket{3',-3'}$ transition. Its propagation direction is precisely aligned with the quantization axis of the atoms, which is jointly defined by the magnetic bias field of 1.5 Gauss and the electric field direction of the linearly polarized optical dipole trap. The circularity of the $\sigma^-$ illumination  light has been experimentally optimized to reduce off-resonant excitations to the states $\ket{2',-2'}$ and $\ket{2', -1'}$, thereby suppressing the dominant population leakage channel from the bright state into the dark manifold during state detection (see insets in Fig.~\ref{fig:fig1}a). The excellent control over the light polarization obtained by using a single illumination beam, however, comes at the cost of losing the 3D Doppler polarization gradient cooling during photon scattering of bright atoms. As a consequence, survival of the atoms in the optical trap decreases for longer state detection intervals, and a careful choice of illumination parameters is necessary to optimize both state detection efficiency and atom survival probability.

\subsubsection{Measurement sequence}\label{sec:ExpSeq}

To characterize the state detection imaging technique, we determine the number of scattered photons and the atom loss probability as a function of the illumination time for atoms in the bright and the dark state. For this purpose the atoms are subjected to the sequence of alternating PI and SI shown in Fig.~\ref{fig:fig1}. \\ \indent
During PI the dipole trap depth is held at $k_{\rm{B}}\times\SI{1.5}{\milli\kelvin}$ and molasses illumination is applied during \SI{20}{\milli\second} with cooling light at an intensity of $3\cdot I_\text{sat}$ intensity and  a detuning of $-4.4\cdot\Gamma$ relative to the unshifted $\ket{2,-2} \rightarrow \ket{3',-3'}$ transition together with repumping light ($F=1 \rightarrow F=2$) with an intensity of about $1\cdot I_{\text{sat}}$ intensity. \\ \indent
For state detection imaging we explore the parameter space by varying the dipole trap depth $U_{0}$ and the SDB detuning $\Delta_{\rm{SDB}}$, intensity $I_{\text{SDB}}$, and illumination time $t_{\text{SDB}}$. The SDB is applied to atoms optically pumped into the bright (SI1 in Fig.~\ref{fig:fig1}b) and into the dark state (SI2 in Fig.~\ref{fig:fig1}b). We also record an image without illumination (SI3 in Fig.~\ref{fig:fig1}b) to characterize atom losses that are not related to the SDB. By comparing the atom positions using PI before and after SI, the atom loss probability related to the illumination process is determined.\\ \indent
The experimental results are presented in Secs.~\ref{sec:ExpMeas1} and~\ref{measurementIntense}. In order to interpret the measurements we first present a physical understanding  of the  dynamics of the atom in the dipole trap illuminated by the SDB.

\section{Heating dynamics of optically trapped atoms under near-resonant illumination}\label{HeatingDynamics}

In standard  methods for laser cooling  of atoms, such as Doppler, Polarization Gradient, Raman, Microwave, EIT and Cavity cooling  schemes, atoms are illuminated by near-resonant fields. The associated population of excited states causes optically-trapped atoms to experience a dipole force frequently opposing the confinement action of the optical lattice. In practice, the cooling schemes are not limited by this effect and it is commonly assumed that the anti-confining potentials are not relevant for the motional dynamics of the atoms during near-resonant illumination~\cite{gibbons2011}. In consequence, their heating contribution has not received  adequate theoretical attention~\cite{taieb1994} since the early days of laser cooling~\cite{Gordon1980,dalibard1985} and only recently it has been considered to improve Raman cooling in an optical lattice~\cite{cheuk2015quantum}. For our state-selective near-resonant illumination conditions with a single SDB, the absence of continuous 3D cooling means that all heating effects must be considered. In the following sections, we first summarize the heating induced by photon recoil (Sec.~\ref{photonrecoilheating}). Then, we introduce a model for the anti-confining excited state potential (Sec.~\ref{Sec:DipoleForceFluct}) and analyze in detail the effect of the dipole force fluctuation for two different regimes: For a weak, resonant SDB (Sec.~\ref{toymodel})  and for an intense, detuned illumination field (Sec.~\ref{sec:offresonantexcitation}). We present the results of Monte Carlo simulations for both heating mechanisms and compare the theoretical models to experimental data .\\ \indent

\subsection{Photon recoil heating}
\label{photonrecoilheating}

To quantify the heating induced by photon recoil during the illumination process, we assume that each scattering event induced by a single, near-resonant light field with wavelength $\lambda_{\rm{Rb}} = {2\pi}/{k_{\rm{Rb}}}$ increases on average the kinetic energy of an atom with mass $m$ by an amount comparable to twice the recoil energy $E_{\rm{rec}} = {\hbar^2k_{\rm{Rb}}^2}/{2m}$. By furthermore making the approximation that an initially cold atom is lost once its average energy exceeds the ground state trap depth $U_0$, it takes in average $N=U_0/2 E_{\rm{rec}}$ photons to lose an atom. This number would simply be proportional to the trap depth and, for instance, correspond to $\approx 4800$ photons for $U_0= k_{\rm{B}} \times \SI{3.46}{\milli\kelvin}$. \\ \indent

\subsection{Dipole force fluctuations}
\label{Sec:DipoleForceFluct}

In addition to photon recoil heating, the random \emph{dipole force fluctuation} (DFF) caused by excitation and decay from the anticonfined exited state couples the atom's internal and external degrees of freedom. We will show that DFF in deep optical potentials can severely reduce the number of scattered photons of the SDB before an atom  is lost from the trap. We take advantage of the simplicity of our experimental setup to quantitatively measure and interpret this heating mechanism.  The goal is to develop models that describe DFF heating induced by photon scattering in optical potentials for parameter regimes that are relevant to fluorescence imaging for a wide range of neutral atom experiments~\footnote{In some experiments the difference in the trapping potentials of ground and optically excited states, and thus the associated DFF heating, can be eliminated by choosing a species-specific \emph{magic wavelength} or using bichromatic optical fields. Many times, however, the restrictions on the choice of the optical trapping potential conflict with other experimental constraints and therefore prevent the application of these techniques.}. \\ \indent

\subsubsection{State-dependent optical dipole trap potential}

We begin our theoretical treatment of an atom that simultaneously interacts with the two relevant light fields by highlighting the different mechanisms with which the dipole trap and the SDB contribute to the dynamics of the atom: The strong, far-detuned dipole trap is responsible for trapping and repulsive forces for the electronic ground and excited state that determine the atomic motion. The homogeneous illumination by the near-resonant {SDB}, on the other hand, mainly gives rise to fluorescent transitions between the internal states of the atom. \\ \indent
The properties of the atoms perturbed by the far-off-resonant light field, including their dipole transitions, still correspond closely to the bare atomic states. We thus approximate the atomic states relevant for photon scattering by the unperturbed \emph{bright} ground state $\ket{g} \equiv \ket{2,-2}$ and optically excited state $\ket{e} \equiv \ket{3',-3'}$ with position-dependent energy shifts as sketched in Fig.~\ref{fig:ExpDataWeakModel}. The dipole trap induced scattering ($< \hspace{-0.2em}0.08$ photons per millisecond) is  much smaller than the scattering rate due to the SDB and it is therefore neglected. Using this approximations, the Hamiltonian for the dipole trapped atom is given by

\begin{equation}\label{eq:AtomInDT}
\mathbf{\hat{H}}_{\text{A-DT}}=\frac{p^{2}}{2m} + \left[\hbar \omega _{0}+U_{e}\left( \mathbf{r}\right) \right] \left\vert  {e}\right\rangle \left\langle   {e}\right\vert +U_{g}\left( \mathbf{r}\right) \left\vert{  {g}}\right\rangle \left\langle {  {g}}\right\vert ,
\end{equation}
where $\omega _{0}$ is the atomic resonance frequency in free space, $U_{g}\left( \mathbf{r}\right)$ and $U_{e}\left( \mathbf{r}\right)$ are the conservative potentials induced by the dipole trap for the ground and excited state, respectively and $U_{g}(0) = U_0$.\\ \indent

Illuminating the atom with the near-resonant SDB induces transitions between the two states and their associated potentials and therefore drives a coupled evolution of the internal and external degrees of freedom. As a result, trapped atoms in a coherent superposition of ground and excited states will experience a splitting of their atomic wave packet. A fully quantum-mechanical treatment of the problem, e.g. using quantum Monte Carlo methods~\cite{plenio1998quantum,dum1992monte,molmer1993monte,Castin1995}, quickly becomes impracticable due to the large number of motional states in the deep, anharmonic, three-dimensional potentials. For two limiting regimes, however, the problem can be approximated using methods that treat the motion of the atoms classically:  \\ \indent 
In the regime of weak, resonant illumination (Sec.~\ref{sec:WeakResonantModel}), where spontaneous scattering dominates, we neglect coherent evolution of the internal atomic state and consider transitions between the ground and excited state potential, and vice versa, as instantaneous photon absorption and emission events. \\
 \indent In the regime of an intense, detuned illumination field (Sec.~\ref{sec:offresonantexcitation}), we describe the interaction with the illumination light using the \emph{dressed state formalism} and consider photon scattering as transitions between dressed states. These cases are discussed in detail below and compared with measurements for a wide range of experimental parameters.
 
\subsection{Heating induced by weak, resonant illumination} \label{sec:WeakResonantModel} 
\label{toymodel}
To illustrate how jumps between different conservative potentials caused by instantaneous absorption and emission of SDB photons lead to heating of a trapped atom (see Fig.~\ref{fig:ExpDataWeakModel}), we first consider a simplified one-dimensional toy model of the process.
  
\subsubsection{Toy model for a one-dimensional harmonic potential}
\begin{figure}[t]
    \includegraphics[width=1.0\columnwidth]{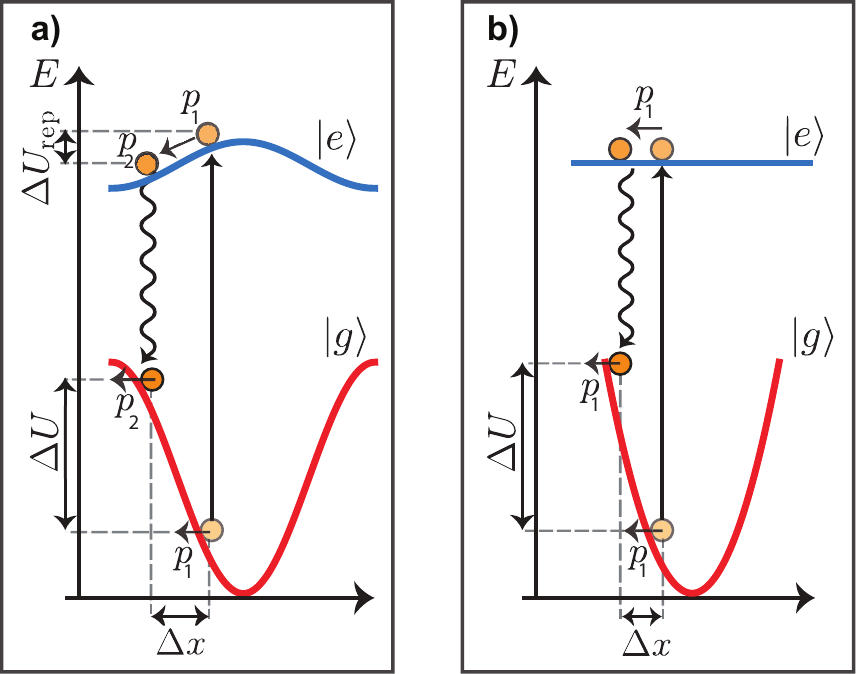}
  	\caption {Dynamics of an optically trapped atom during fluorescent scattering according to the picture of absorption and emission jump events. \textbf{a)} Photon scattering is assumed to give rise to instantaneous transitions of the atom between the trapping potentials of the ground and excited state. \textbf{b)} Simplified potential shapes of the toy model.} {\label{fig:ExpDataWeakModel}}
\end{figure}

In order to obtain  an analytic solution, the toy model assumes a flat excited state potential (see Fig.~\ref{fig:ExpDataWeakModel}b). We start by considering an atom in the ground state at position $x$ in the trap with momentum $p$ and total energy $E=U_{g}\left( x\right) +{p^{2}}/{2m}$. After an excitation from the ground state, the atom remains in the excited state for a time $t$, where in the absence of a confining potential it travels at constant velocity for a distance $\Delta x=\frac{p}{m}t$. Once it decays back to the ground state, the energy change due to the displacement is given by 
$\Delta E =U_{g}\left( x+\Delta x\right) -U_{g}\left( x\right)$.\\ \indent

By Taylor-expanding the ground state potential and by assuming an exponentially-distributed time in the exited state (with time constant $\Gamma^{-1}$), we obtain 
\begin{equation}\label{eq:EnergyCahnge2}
\left\langle \Delta E\right\rangle _{t}=\sum_{n=1}^{\infty } U_{g}^{^{\left(n\right) }}\hspace{-0.5em}\left( x\right) \left( \frac{p}{\Gamma m}\right) ^{n}
\end{equation}
for the mean energy change per scattering event at position $x$, where $\langle\cdot\rangle_{t}$ denotes the average over time and $^{\left(n\right)}$ indicates the $n^{\rm{th}}$ derivative.\\ \indent
In a one-dimensional conservative potential every point $x$ in the trap is crossed equally often in the forward and backward direction, i.e. with positive and negative momentum. Therefore averaging Eq.~(\ref{eq:EnergyCahnge2}) over time the odd terms in the expansion cancel and only the even powers in momentum remain:
\begin{equation}\label{eq:EnergyCahnge3}
\left\langle \Delta E\right\rangle _{t}=\sum_{n=1}^{\infty } U_{g}^{^{\left(2n\right)}}\hspace{-0.5em}\left( x\right) \left( \frac{p}{\Gamma m}\right) ^{2n}.
\end{equation}
Assuming small displacements, i.e. excited state lifetimes that are short compared to the trap oscillation time, the first term in Eq.~(\ref{eq:EnergyCahnge3}) dominates for many practical configurations of optical traps, including standing-wave potentials, and we can write
\begin{equation}\label{eq:EnergyCahnge4}
\left\langle \Delta E\right\rangle _{t}\approx U_{g}^{\prime \prime }\left( x\right)
\left( \frac{p}{\Gamma m}\right) ^{2}.
\end{equation}
This means that the average net energy change induced due to DFF is caused by the curvature of the trapping potential and that positive and negative curvatures lead to heating and cooling, respectively. DFF heating is strongest along the axis of the tightly confining, optical, standing-wave potentials, whereas weaker effects are expected along the radial direction of focused Gaussian beams associated with lower trap frequencies.\\ \indent

By approximating the ground state potential of the atom as a purely harmonic potential ($U_{g}(x) = U_{\rm{const}}^{\prime \prime}\cdot x^2/2$), we can furthermore calculate the energy change per scattering (i.e. combined absorption and emission) event 
\begin{equation}\label{eq:EnergyCahnge5}
\left\langle \Delta E\right\rangle _{t,x} =\int \rho _{\text{HO}}(x,E)\left\langle \Delta E\right\rangle _{t}\left( x\right)\text{d}x=\frac{2 U_{\rm{const}}^{\prime \prime}E}{m\Gamma ^{2}},
\end{equation}
where we have averaged over the position probability distribution $\rho_{\text{HO}}(x,E)=1/\pi \sqrt{x_\text{max}^2-x^2}$ of the harmonic oscillator. $x_\text{max}$ is the position of the turning point of the atom with total energy $E=U_{g}\left( x\right) +{p^{2}}/{2m}$~\cite{robinett1995quantum}. According to Eq.~(\ref{eq:EnergyCahnge5}) pure DFF heating would give rise to an exponential energy gain of the atom during fluorescent scattering. \\ \indent
To compare heating due to DFF with the additional energy-independent heating effects caused by photon recoil, we consider an atom trapped in a standing wave optical potential ($U_{\rm{const}}^{\prime \prime} \approx  -U_0 k_{\rm{DT}}^2$ for low energies) with optical depth $U_0$ and dipole trap laser wavelength $\lambda_{\rm{DT}}=2\pi/k_{\rm{DT}}$. 
We find that, for the simplified potential shapes in our model, the energy increase per scattered photon due to DFF 
\begin{equation}\label{eq:EnergyChangeCompRecoil}
\frac{\left\langle \Delta E\right\rangle _{t,x}}{E_\text{rec}}=\frac{4\lambda _{\text{Rb}}^{2}U_{0}E}{\lambda _{\text{DT}}^{2}\Gamma ^{2}\hbar ^{2}},
\end{equation}
can become larger than photon recoil heating rate for trap depths that exceed 
\begin{equation}\label{eq:300muK}
U_0>\frac{\lambda _{\text{DT}} \Gamma \hbar}{2\lambda _{\text{Rb}}}\approx k_\text{B} \times \SI{300}{\micro\kelvin}.
\end{equation}
The analytic toy model presented here provides an intuitive understanding of the DFF heating mechanism, but does neither consider the repulsive potential for the excited state $U_{e}$, nor the strong anharmonicity of the potentials. These two effects, which are expected to lead to additional heating and cooling contributions, respectively, are considered in the Monte Carlo treatment in Sec.~\ref{MonteCarlo}.

\subsubsection{Monte Carlo simulation of heating}
\label{MonteCarlo}

In order to perform a quantitative comparison with our experimental data, we have carried out  a simulation of the heating effects during fluorescent scattering, which accurately represents the experimental system, including the anharmonic, three-dimensional nature of the optical potentials.
Early, pioneering studies of the dynamics of optically trapped atoms by Gordon and Ashkin~\cite{Gordon1980} and Dalibard and Cohen-Tannoudji~\cite{dalibard1985} determined momentum diffusion coefficients from semi-classical models to investigate the motional dynamics in state dependent potential using Fokker-Planck equations (FPEs). With coefficients and FPEs for the total energy, atom loss due to heating in conservative optical traps was also studied (under the assumption of harmonic confinement) by calculating the evolution until the energy of the particle exceeds the trap depth~\cite{gehm1998}. \\ \indent
In the case of optically trapped atoms illuminated with near-resonant light, numerical simulations of the atom dynamics based on FPEs are complicated by the strongly position-dependent light shifts in deep optical potentials. The position-dependent shifts of the effective detuning for the SDB modify the local photon scattering rates and thereby couple the motion degrees of freedom along the three trap axis. For this reason we have implemented a classical Monte Carlo simulation (MC) to reproduce the coupled evolution of the internal and external degrees of freedom due to photon scattering. 

The trajectory-based approach of a MC calculation moreover makes it easier to include additional scattering channels of the multilevel atom, which can lead to the transfer of the atom into the \emph{dark} hyperfine manifold. \\ \indent

The MC simulation for the \emph{DFF model} implements the full heating dynamics in the picture of absorption and emission jumps by numerically solving the equations-of-motion in the confining and anti-confining potential, for atoms in the ground and excited state, respectively. For an atom in the ground state, the next photon absorption event is randomly chosen according to the position- and velocity-dependent~\footnote{Effects due to Doppler-shifts are thus included in the Monte Carlo simulation, but turn out to play no significant role for the heating and cooling dynamics during fluorescent scattering in the optical potentials.} instantaneous scattering rate (see Appendix~\ref{app:MCSimulation}). Each absorbed photon, adds one recoil momentum to the atomic motion along the propagation direction of the SDB. Following photon absorption the atom remains in the excited state for a time $t_\text{exc}$, which is randomly chosen from the exponential distribution $\rho(t)=\Gamma\exp(-\Gamma t)$ before it decays back to the ground state. Each spontaneous emission adds one recoil momentum along a direction that is randomly chosen according to the dipole radiation distribution of sigma polarized light. Further details regarding the simulation, including the modeling of transitions to the dark state are provided in Appendix~\ref{app:MCSimulation}. \\ \indent
For comparison with the experimental data we also simulate the \emph{recoil model} neglecting all DFF effects, i.e. only considering energy changes by photon recoil (cf. Sec.~\ref{photonrecoilheating}). The MC simulation for the \emph{recoil model} ignores the excited state potential and calculates the trajectory in between scattering events as if the atom remained in the ground state potential $U_g$ throughout. The effects due to photon recoil and the random sampling of photon absorption and emission times are treated identically to the \emph{DFF model}.\\

\begin{figure}[t]
    \includegraphics[width=1.0\columnwidth]
{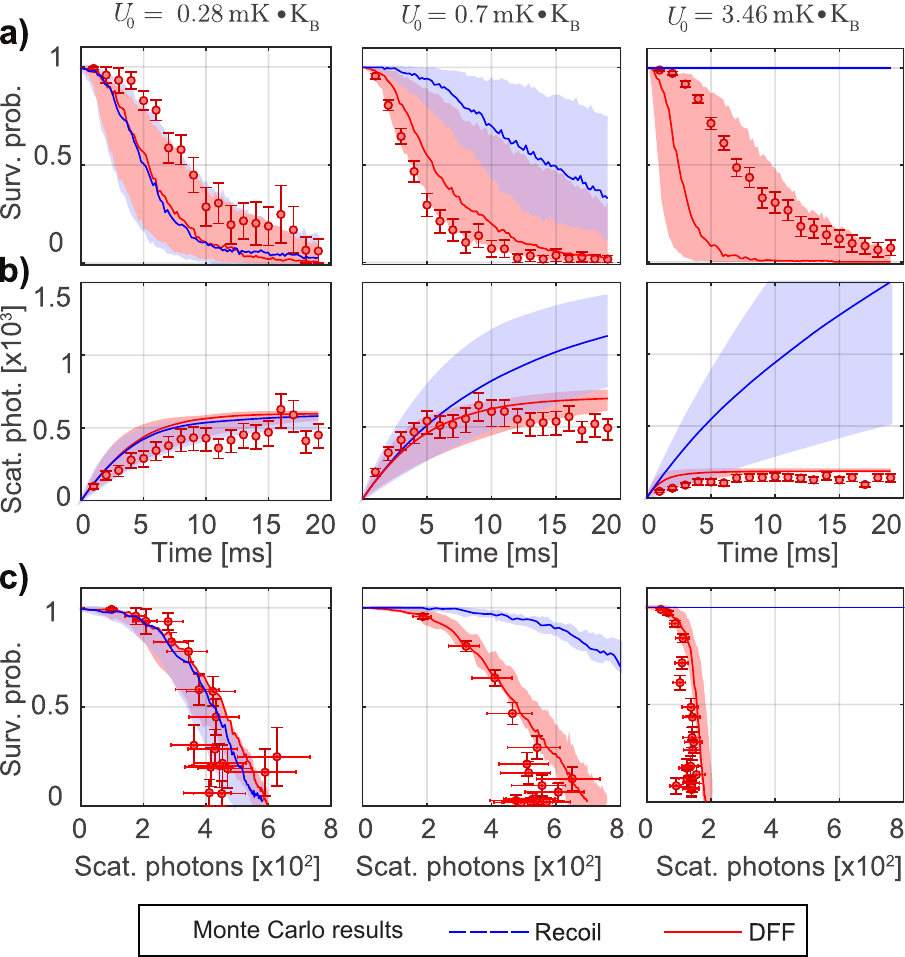}
  	\caption{Dynamics under weak, resonant illumination for different depths of the optical trap. \textbf{a)}~Survival probability and \textbf{b)}~number of scattered photons as a function of the illumination time for an atom initially trapped in the \emph{bright} state. The simulation results obtained for the best-estimate values of the independently characterized experiment parameters are indicated by the \emph{blue dashed} and \emph{red solid} lines for the \emph{recoil} and the \emph{DFF model}, repectively. The shaded region indicated the uncertainty of the simulated results (see text for details). \textbf{c)}~The survival probability for a given number of scattered photons, which is a more important figure of merit for non-destructive, state-selective fluorescence imaging. Error bars of the experimental data in this and all further plots of the article indicate 95\% confidence intervals.} {\label{fig:ExpDataWeakNRF}}
\end{figure}

\subsubsection{Measurements of heating vs. trap depth} \label{sec:ExpMeas1}
The measurements to study the heating induced by DFF in the weak resonant excitation regime are performed using experimental sequences similar to the ones described in Sec.~\ref{sec:ExpSeq} and Fig.~\ref{fig:fig1}. For three depths of the dipole trap $U_{0,\text{meas}} = k_{\rm{B}} \times \{0.28, 0.7, 3.46\}\,$mK, atoms in the \emph{bright} state are illuminated by the SDB with detunings (relative to the unshifted atomic resonance) of $\Delta_\text{SDB,meas} = 2\pi\times \{6,~15,~79\}\,$MHz, respectively. In this way the SDB is approximately resonant with the AC-light shifted $\ket{g}\rightarrow\ket{e}$ transitions at the points of maximal energy splitting ($x=0$), for all three trap depth, respectively. The intensity of the homogeneous illumination by the SDB is set to $0.015\cdot I_{\text{sat}}$.\\ \indent  
At each trap depth we determine for various illumination times the number of scattered photons during SI (using the recorded number of photons and the independently measured photon detection efficiency of Appendix A) and extract the survival probabilities considering the preceding and subsequent PI (cf. Fig.~\ref{fig:fig1}). The measurements are shown in Fig.~\ref{fig:ExpDataWeakNRF}a-c together with the results of the two different theoretical models explained below.\\ \indent

\subsubsection{Comparison of experiment and theory}

In Fig.~\ref{fig:ExpDataWeakNRF} the number of scattered photons and the survival probability for weak resonant illumination are compared with the results of Monte Carlo simulations of the \emph{recoil model} and \emph{DFF model}. For each of the models we perform the simulation using the best-estimate values of the independently measured trap depths $U_{0,\text{meas}}$, illumination intensity $I_\text{meas}$ and detunings $\Delta_\text{meas}$. The shaded bands indicate the uncertainties in the simulation when considering the uncertainties of the experimental parameters ($U_{0,\text{sim}}= U_{0,\text{meas}}\pm 4\%$, $I_\text{sim}= I_\text{meas}\pm 20\%$, $\Delta_\text{SDB,sim}=\Delta_\text{SDB,meas} \pm 1\,$MHz). Assuming adiabatic ramping of the trap potentials from the initial molasses configuration, the initial temperatures of the atoms used for the calculations are $T= \{30, 50, 110\},\mu$K, for the shallow, intermediate and deep trap, respectively. 
The comparison in Fig.~\ref{fig:ExpDataWeakNRF}a and b shows that the predictions of survival probability and the number of scattered photons as a function of time are sensitive to small changes in the experimental conditions. Changes in the detuning and the intensity strongly modify the rate at which the atoms scatter photons. A more important figure of merit is, however, the total gain of energy per scattered photon. This can be visualized when plotting the survival probability as a function of the number of scattered photons (Fig.~\ref{fig:ExpDataWeakNRF}c). This quantity is insensitive to small experimental uncertainties. At the lowest value for the trap depth ($U_{0,\text{meas}}=k_{\rm{B}}\times0.28\,$mK) DFF heating does not play a significant role compared to photon recoil heating (cf. Eq.~\ref{eq:300muK}) and both models (recoil and DFF) yield similar results and agree with experimental data. With stronger confinement of the atoms  ($U_{0,\text{meas}}= k_{\rm{B}}\times 0.7$\,mK) DFF heating starts to dominate over recoil heating. The total number of photons scattered before an atom is lost decreases despite the large increase in the trap depth of the optical potential ($U_{0,\text{meas}}= k_{\rm{B}}\times 3.46$\,mK). \\ \indent
For \emph{weak resonant} illumination we have thus found (for our particular optical trap geometry) a strict limit of a few hundred fluorescence photons that can be scattered before an atom is lost. 
\subsection{Heating induced by intense, detuned illumination} 
\label{sec:offresonantexcitation}
  	
We now investigate the scattering of light from atoms in optical trap potentials using larger detunings for the SDB and higher intensity. For intense, non-resonant illumination, the coherences in the evolution of the atom, which are not captured in the bare-state absorption and emission picture, cannot be neglected anymore.

To intrinsically include the coherent atom-field coupling into our model for the motional dynamics of the system, we make use of the dressed state formalism~\cite{cohen1977dressed,cohen1992atom}  following a similar approach as in Ref.~\cite{cheuk2015quantum}. We start with
\begin{align}
 \label{eq:HamNRL}
\mathbf{\hat{H}} &=\mathbf{\hat{H}}_{\text{A-DT}}+\mathbf{\hat{H}}_{\text{SDB}}+\mathbf{\hat{H}}_{\text{A-SDB}}, \\
\mathbf{\hat{H}}_{\text{SDB}} &=\hbar \omega _{\text{SDB}}\left( \hat{a}_{\text{SDB}}^{\dagger }\hat{a}_{\text{SDB}}\right),\nonumber\\
\mathbf{\hat{H}}_{\text{A-SDB}} &=\frac{\hbar \Omega _{0,\text{SDB}}}{2}\left( \hat{\sigma}^{\dagger }\hat{a}_{\text{SDB}}+\hat{\sigma}\hat{a}_{\text{SDB}}^{\dagger }\right)\nonumber,
\end{align}
where we consider the coupling with the non-resonant SDB in the rotating wave approximation and treat the effect of the dipole trap again as position-dependent AC-Stark shifts to the atomic transitions (cf. Eq.~\ref{eq:AtomInDT}). $\hat{a}_{\text{SDB}}^{\dagger }$,$\ \hat{a}_{\text{SDB}}$ are the creation and annihilation operators, $\Omega _{0,\text{SDB}}$ and $\omega _{\text{SDB}}\,$ denote the resonant Rabi frequency and the angular frequency of the SDB, respectively. \\ \indent
\subsubsection{Near-resonant light  dressed state potentials}

\begin{figure}[t]
    \includegraphics[width=1.0\columnwidth]{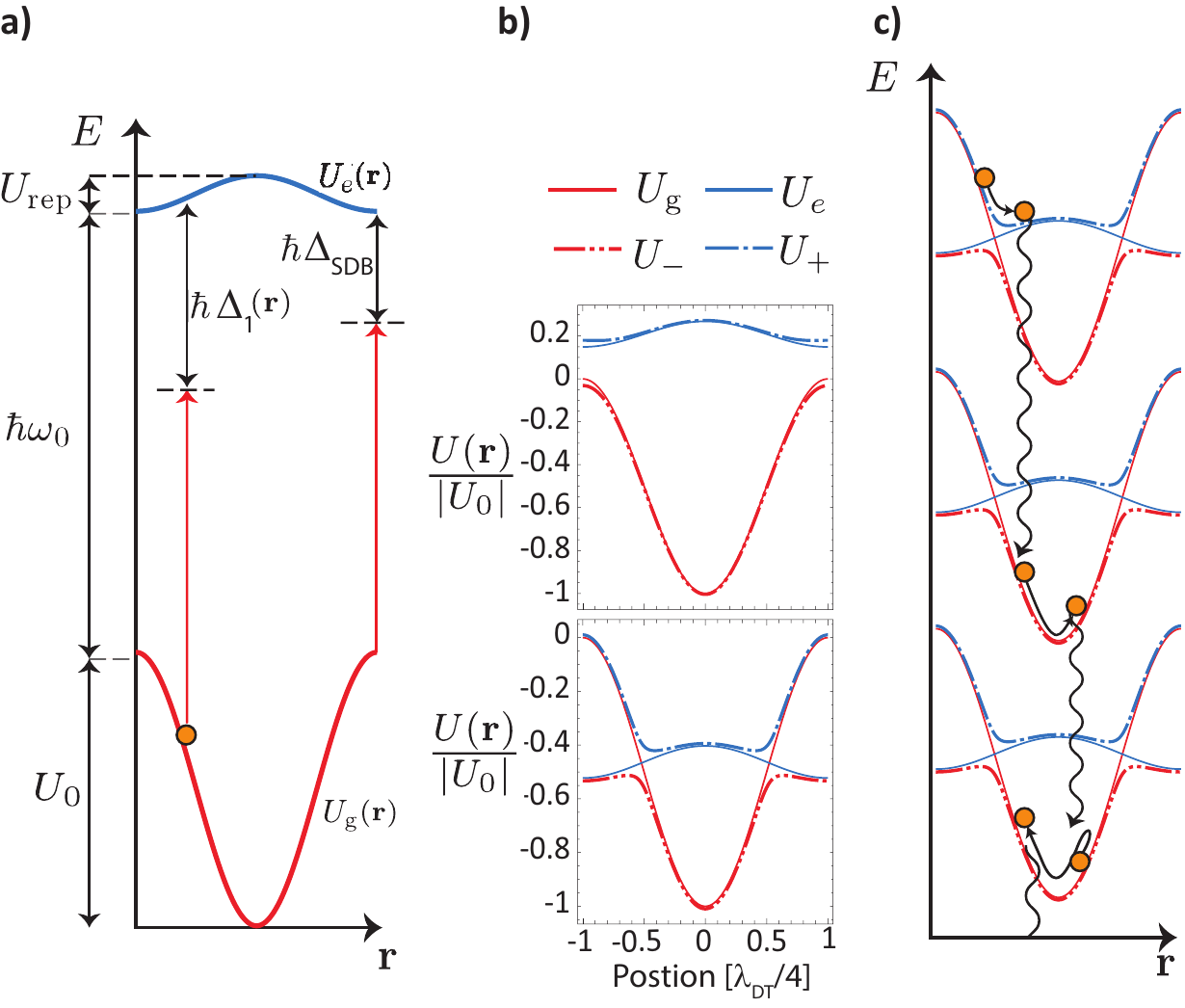}
  	\caption{Dressing of the states of an optically trapped atom by intense, detuned near-resonant illumination. \textbf{a)}~Two-level atom in the dipole trap illuminated by the homogeneous SDB. \textbf{b)}~Examples of dressed states potentials with $U_{0}=3.46\,$mK and $\Omega_\text{0,{SDB}}/2\pi=\SI{35}{\mega\hertz}$ for $\Delta_\text{SDB}/2\pi=\SI{-5}{\mega\hertz}$ (top), and $\Delta_\text{SDB}/2\pi=\SI{+39}{\mega\hertz}$ (bottom). For the first case, the atom is never in resonance with the SDB  and for the second, the atom is in resonance with the SDB at $\mathbf{r}_0\approx \pm0.5\lambda_\text{DT}/4$. \textbf{c)} Photon scattering in the dressed-state picture. An atom trapped in a dressed state potential transitions to another dressed state potential by removing one photon from the SDB and emitting it into free space.} {\label{fig:DressedStates}}\end{figure}
  	\begin{figure*}[t] 
    \includegraphics[width=1 \textwidth]{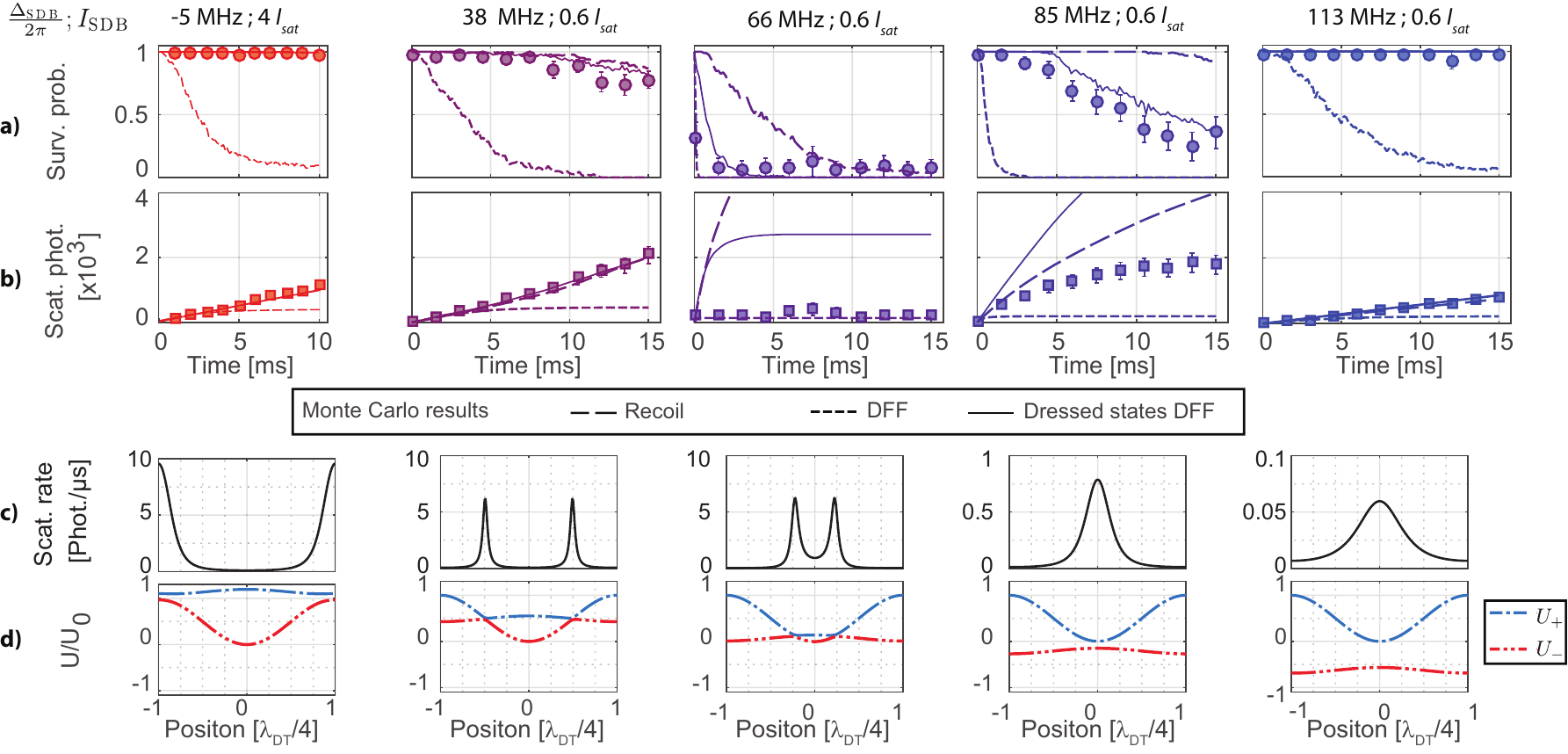}
  	\caption{Fluorescence dynamics under intense near-resonant illumination for different detunings and intensities. \textbf{a)}~Survival probability and \textbf{b)} number of scattered photons as a function of the illumination time for an atom initially trapped in the \emph{bright} state. The Monte Carlo simulation results for the \emph{dressed states DFF model} are shown together with the pure photon \emph{recoil model} and dipole force fluctuations \emph{DFF model} described in Sec.~\ref{MonteCarlo} (cf. Fig.~\ref{fig:ExpDataWeakNRF}). \textbf{c)} The position dependent photon scattering rate and \textbf{d)} the dressed states potentials $U_{-}(\mathbf{r}=(x,0,0))$ and $U_{-}(\mathbf{r}=(x,0,0))$ are displayed for positions along the standing-wave trap axis. The trap depth is $k_{\rm{B}}\times 3.46$\,mK.} \label{fig:ExpDataResonance}
\end{figure*}
The eigenstates of the Hamiltonian in Eq.~(\ref{eq:HamNRL}) are the SDB-dressed states
\begin{eqnarray}\label{eq:NRL_DressedStates}
\left\vert +,N _{\text{SDB}}\right\rangle &=&\sin(\theta\left(\mathbf{r}\right))\left\vert   {g},N _{\text{SDB}}\right\rangle\\ \nonumber
&&+\cos(\theta\left( \mathbf{r}\right) )\left\vert   {e},N _{\text{SDB}}-1\right\rangle,  \\
\left\vert -,N _{\text{SDB}}\right\rangle &=&\cos(\theta\left(\mathbf{r}\right)) \left\vert   {g},N _{\text{SDB}}\right\rangle \\
&&-\sin(\theta\left( \mathbf{r}\right)) \left\vert   {e},N _{\text{SDB}}-1\right\rangle, \nonumber  
\end{eqnarray}
where the mixing angle is defined by
\begin{eqnarray}\label{eq:MixingAngleNRF}
\theta &=& \frac{1}{2}{\rm arctan}\Big(-\frac{\Omega_\text{0,SDB}}{\Delta_1}\Big) + \frac{\pi}{2}H(\Delta_1),\\
\Delta _{1}\left(\mathbf{r}\right)  &=&\Delta_{\text{SDB}}+\frac{U_{g}\left(\mathbf{r}\right) - U_{e}\left( \mathbf{r}\right) }{\hbar }.\nonumber
\end{eqnarray} 
Here, $H()$ denotes the Heaviside step function, $\Delta _{\text{SDB}}=\omega _{\text{SDB}}-\omega _{0}$ is the detuning of the SDB from the atomic transition of the untrapped atom and $\Delta _{1}\left(\mathbf{r}\right)$ represents the total detuning of the SDB at position $\mathbf{r}$, which takes into account the AC-Stark shift induced by the dipole trap (see Fig.~\ref{fig:DressedStates}a). The eigenenergies corresponding to the new dressed states in Eq.~(\ref{eq:NRL_DressedStates}) are

\begin{equation}\label{eq:DSEnergies}
E_{\pm ,\text{SDB}} = N_\text{SDB}\hbar \omega _{\text{SDB}} + U_{\pm }\left( \mathbf{r}\right)
\end{equation}
with
\begin{equation}\label{eq:DSPotentials}
U_{\pm }\left( \mathbf{r}\right)=U_{g}\left( \mathbf{r}\right) +\frac{\hbar }{2}\left( -\Delta _{1}\left( \mathbf{r}\right) \pm \Omega _{_{\text{SDB}}}\left( \mathbf{r}\right) \right), 
\end{equation}

where $\Omega_{\text{SDB}}\left( \mathbf{r}\right)=\sqrt{\Delta _{\text{1}}^{2}\left( \mathbf{r}\right) +\Omega _{0,\text{SDB}}^{2}}$ is the generalized Rabi frequency. \\ \indent

For large (red or blue) detunings ($|\Delta_{\text{1}}({\bf r}) |\gg \Omega_{0,\text{SDB}}$) of the SDB from the resonance, the potentials for the SDB-dressed states $U_{\pm }\left( \mathbf{r}\right)$ are almost identical in shape to the original dipole trap potentials, whereas at resonance ($\Delta_{\text{1}}(\mathbf{r}_0)=0$) the curvature of the dressed state potentials (see Fig.~\ref{fig:DressedStates}) is modified by the occurrence of anti-crossings.

\subsubsection{Photon scattering without change of potentials}

\begin{figure*}[t]
		\includegraphics[width=0.9\textwidth]{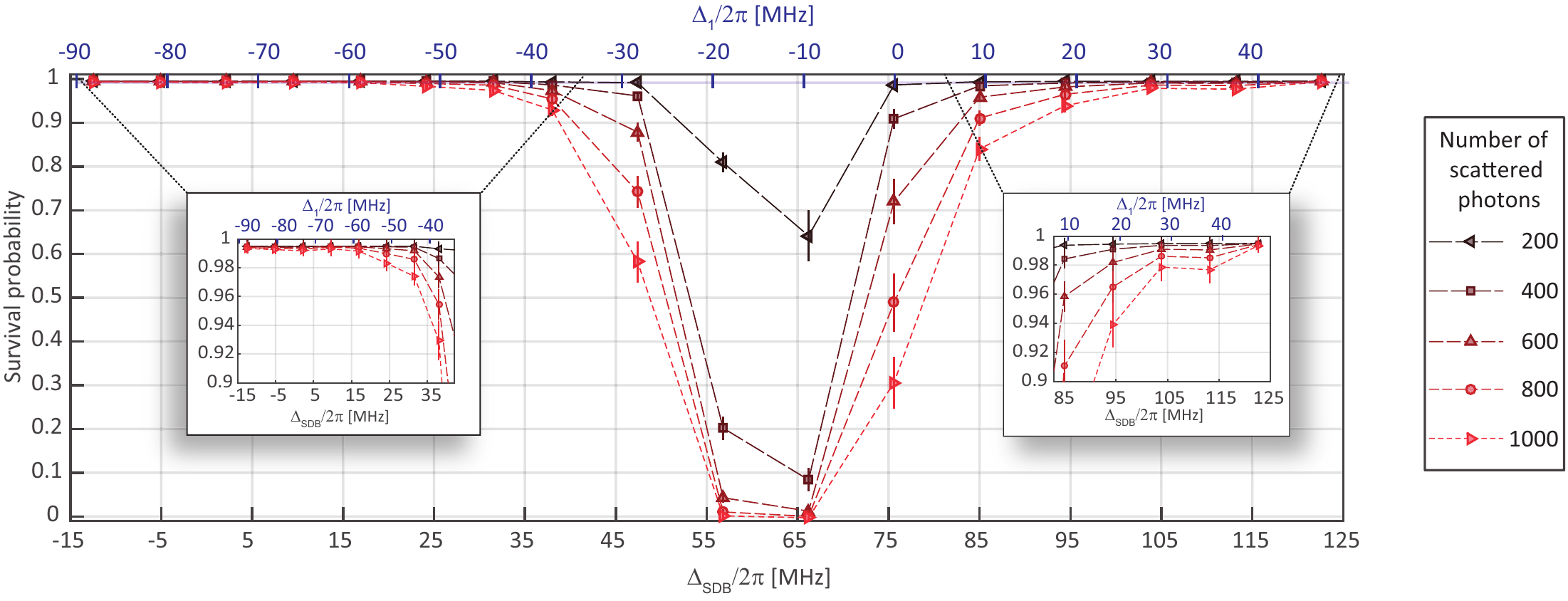}
  	 	\caption{Survival probability of atoms in the trap vs. detuning of the SDB for a given number of scattered photons. The x-axis shows the detuning for free space (bottom) and for the AC-Stark shifted atoms at the minimum of the trap (top blue). The results computed from the data shown in Fig.~\ref{fig:ExpDataResonance} and the full data set described in Sec.~\ref{measurementIntense} show that non-destructive, state-selective fluorescence imaging is best achieved for large blue or red detunings of the SDB. The curves are not centered at the AC-Stark shifted resonance since the atoms spend most of the time not at bottom of the trap. Therefore, for $\Delta_1<0$, the atoms become resonant with the SBD, which however, does not occur for $\Delta_1>0$. The lines connecting the data points serve as guides to the eye. The small insets are a zoom into the data points for red (left) and blue (right) detuning.}\label{fig:ExpDataResonanceAll}
\end{figure*}
The full dynamics between the SDB-dressed-states (DS) due to photon scattering is described by the Optical Bloch Equations (OBE). In the \emph{secular approximation} that applies for large detunings or intense fields ($\Omega_{0,\text{SDB}},|\Delta_{\text{1}}({\bf r})| \gg \Gamma$) the OBE in the dressed state basis can be can be written as~\cite{dalibard1985}
\begin{align}\label{eq:OBESDS}
\dot{\rho}_{++} &=-\Gamma _{+-}\rho _{++}+\Gamma _{--}\rho _{--},\nonumber \\
\dot{\rho}_{--} &=-\Gamma _{--}\rho _{--}+\Gamma _{+-}\rho _{++}, \nonumber \\
\dot{\rho}_{+-} &=-\left( i\Omega _{0,\text{SDB}}+\Gamma _{\text{coh}}\right)
\rho _{+-} ,
\end{align}
where $\rho_{ij}=\sum_{N}\left\langle i,N_{\text{SDB}}\right\vert \rho \left\vert j,N_{\text{SDB}}\right\rangle$, $i\in \left\{ +,-\right\}$ are the reduced populations and coherences. The reduced coherences decay towards their vanishing steady state value at the decay rate  $\Gamma _{\text{coh}} =\Gamma \left( \frac{1}{2}+\cos ^{2}\theta\sin ^{2}\theta\right)$.
Eq.~(\ref{eq:OBESDS}) neglects non-adiabatic transitions induced by the motion of the trapped atoms in the position dependent DS potential~\cite{dalibard1985}, which is a valid approximation for the case ($\Delta_{1}({\bf r}) \gg \Omega_{0,\text{SDB}}$). \\  \indent
The transition rates between the dressed states are given by~\cite{dalibard1985}
\begin{align} \label{eq:RatesDS}
\Gamma _{++}& =\Gamma \sin ^{2}\theta\cos ^{2}\theta,  \ \ \Gamma _{-+} =\Gamma \sin ^{4}\theta,\\
\Gamma _{+-}& =\Gamma \cos ^{4}\theta,  \ \ \ \ \ \ \ \ \ \   \Gamma _{--} =\Gamma \sin ^{2}\theta\cos ^{2}\theta\nonumber.
\end{align}
In the case of red detuning ($\Delta_1({\bf r})<$0) the transition rates evaluate to 
\begin{align}
\Gamma _{++}&  \approx \Gamma \left( \frac{\Omega _{0,\text{SDB}}}{2\Delta _{\text{1}}\left( \mathbf{r}\right) }\right) ^{2}, \ \ \ \  \Gamma _{+-} \approx \Gamma \\
\Gamma _{-+} &\approx\Gamma \left( \frac{\Omega _{0,\text{SDB}}}{2\Delta _{\text{1}}\left( \mathbf{r}\right) }\right) ^{4}, \ \ \ \ \Gamma _{--} \approx \Gamma \left( \frac{\Omega _{0,\text{SDB}}}{2\Delta_{\text{1}}\left( \mathbf{r}\right) }\right)^{2}\nonumber.
\end{align}
Due to the relative strength of the transition rates, atoms most strongly populate the state $\left\vert -,N_{\text{SDB}}\right\rangle $ and most likely decay $\left\vert -,N_{\text{SDB}}\right\rangle  \rightarrow \left\vert -,N_{\text{SDB}}-1\right\rangle $ during steady state photon scattering. As a consequence, for most photon scattering events the atom is confined by the same trapping potential ($U_-$) and hence no energy change due to DFF occurs (see Fig.~\ref{fig:DressedStates}). In analogy, photon scattering happens predominantly between ($U_+$) potentials for blue detuning ($\Delta_1({\bf r})>$0).

\subsubsection{Measurements of heating vs. detuning}
\label{measurementIntense}

The experimental setup and sequence to explore the heating for different illumination conditions are described in Sec.~\ref{sec:ExpSeq}. Using a trap depth of $k_{\rm{B}}\times 3.46$\,mK, the survival and the number of scattered photons were measured as a function of the illumination time for different sets of SDB parameters. For SDB detunings in the range $\Delta_{\text{SDB}}/2\pi=+38{\,}$MHz and $+$112${\,}$MHz data was taken at a SDB intensity of 0.6${\,}I_{\text{sat}}$, while for the larger detunings in the range $\Delta_{\text{SDB}}/2\pi=-12\,$MHz to $+39\,$MHz and at $+123\,$MHz the data was recorded at intensities of $4\,I_{\text{sat}}$ and $1.9\,I_{\text{sat}}$, respectively. For a few selected detunings the time-dependent survival and photon scattering curves are shown in Fig.~\ref{fig:ExpDataResonance}. The full set of recorded data has been used to compute the results in Fig.~\ref{fig:ExpDataResonanceAll}, which shows the survival probabilities for a given number of scattered photons, as a function of the SDB detuning. This quantity, which represents a relevant figure of merit for non-destructive, state-selective fluorescence imaging, is clearly optimized for large blue or red detunings of the SDB.\\ \indent

\subsubsection{Comparison experiment and theory}
The measured, time-dependent survival and photon scattering curves in Fig.~\ref{fig:ExpDataResonance} are plotted with the theoretical results of Monte Carlo simulation from three different heating models. The first two models are the \emph{recoil model} and the \emph{DFF model} introduced in Sec.~\ref{MonteCarlo}. In the third model (\emph{dressed state DFF model}) the three-dimensional equation of motion is solved for the atoms in the DS potentials of Eq.~(\ref{eq:DSPotentials}). The jumps events between the DS potentials are chosen in the MC simulation according to the population changing rates in Eq.~({\ref{eq:RatesDS}}). 
The parameters assumed in the simulations have been adjusted within their experimental measurement uncertainties (cf. Sec.~\ref{Appendix}) to $I_{\text{sim}}=0.815\,I_{\text{exp}}$,  $\Delta_{\text{SDB,sim}}=\Delta_{\text{SDB,exp}}+0.5$~MHz, $U_{0,\text{sim}}=0.96\,U_{0,\text{exp}}$ and $T=\SI{140}{\micro\kelvin}$ to improve the fit of the \emph{recoil} and \emph{dressed state DFF} model to the data with large blue or red detuning. For illumination settings ($\Delta_\text{SDB}=\{-5,38,113\}\,$MHz in Fig.~\ref{fig:ExpDataResonance}), where atoms scatter most of the photons with large red or blue effective detunings ($\Delta_1({\bf r})$) the \emph{dressed states DFF model} agrees well with the measurement data and predicts heating dynamics that are dominated by the \emph{recoil} effect. At smaller detunings ($\Delta_\text{SDB}=85\,$MHz in Fig.~\ref{fig:ExpDataResonance}), the \emph{dressed states DFF model} predicts increased heating by dipole force fluctuations from transitions between dressed state potentials and continues to agree better with the measured data than the \emph{recoil} or the \emph{DFF model}. For weak resonant excitation ($\Delta_\text{SDB}=66\,$MHz in Fig.~\ref{fig:ExpDataResonance}) the secular approximation used in the \emph{dressed state DFF model} fails and it underestimates the heating effect. This regime is well decribed by the \emph{DFF model} of Sec.~\ref{sec:WeakResonantModel}.\\ \indent
We conclude that heating process for atoms in standing-wave optical potentials due to photon scattering can be well described in two different regimes by simple classical models. In the intermediate regime, however, a full quantitative description would have to consider the effect of coherences of atomic states and non-adiabatic transitions due to Landau-Zener crossing, which are neglected in our models. 
\section{State-dependent fluorescence imaging}\label{StateDependentImaging}
\begin{figure}[t]
    \includegraphics[width=1.0\columnwidth]{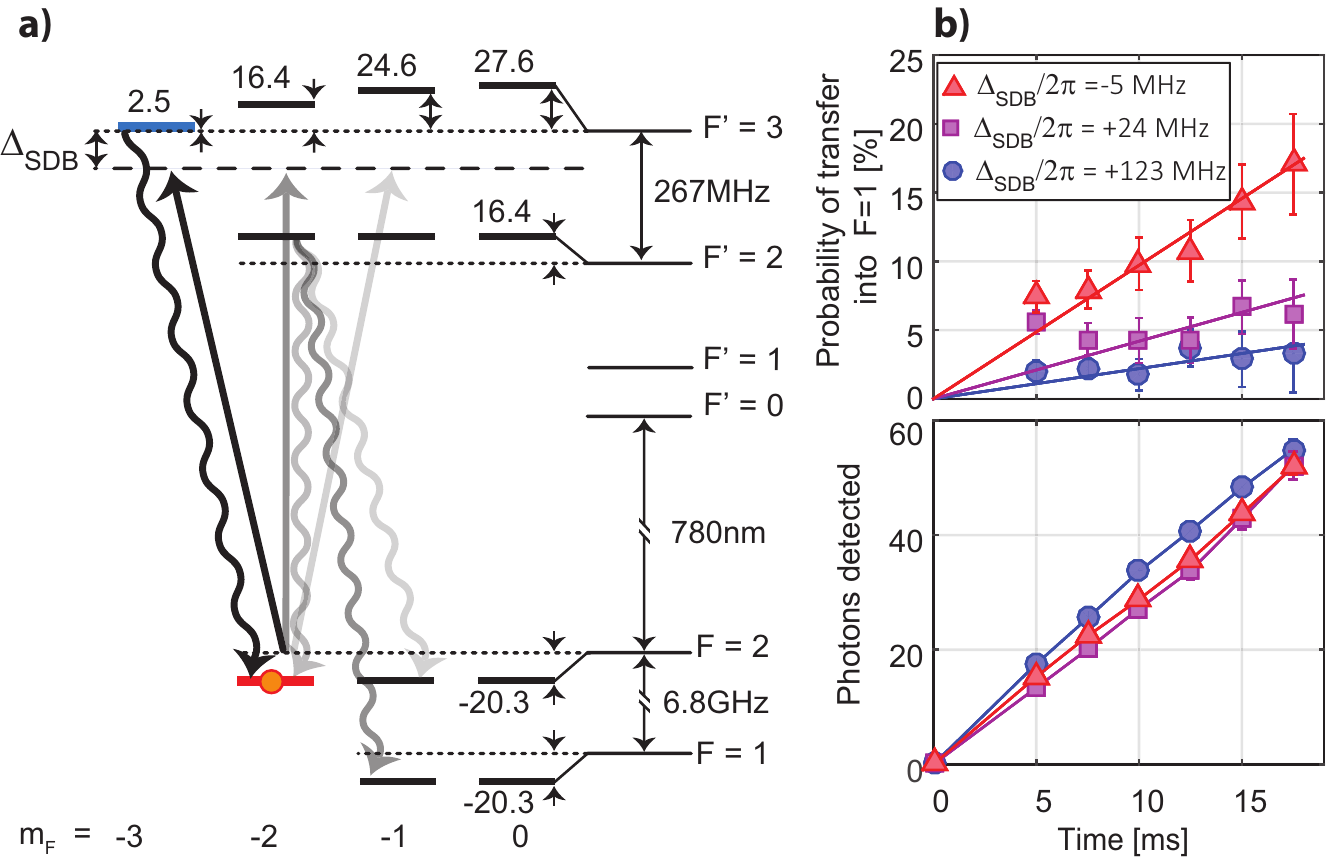}
  	\caption{Leakage into the (\emph{dark}) F=1 state manifold. \textbf{a)}~AC-Stark shifts for a linear $\pi$-polarized dipole trap at 860$\,$nm wavelength interacting with $^{87}$Rb. The numbers in the level diagram indicate the shifts in units of MHz/mK trap depth. Only Zeeman levels relevant to the fluorescent scattering are shown. Leakage into the \emph{dark} state occurs by spontaneous decay, subsequent to excitation into the F'=2 manifold by predominantly $\pi$-polarized contamination photons (see Appendix). \textbf{b}) Atoms transferred to the state $ F = 1$ (top) and number of detected photons (bottom) for different detunings and illumination times. The lines represent a linear fit to the data (top) and a guide to the eye (bottom). All error bars represent 95\% confidence intervals.
} {\label{fig:OffResonantScatt}}
\end{figure}
Fluorescence-based state detection as described in Sec.~\ref{introFluorescence}, relies on distinct photon count signals for atoms that are initially in the bright or the dark state. In previous sections we have shown, that for atoms in strongly confining optical traps, high survival probabilities and a large number of scattered photon can only be achieved by suppressing DFF heating with either blue or red detuned illumination light. 

\subsection{Leakage to the dark state}

Whereas both red and blue detunings lead to equal amounts of energy transfer per scattered photon, the choice of detuning gives rise to different leakage rates between the \emph{bright} and the \emph{dark} states. In order to achieve the low leakage rates to the dark state, that are required for high-fidelity fluorescence state detection, the closed cycling transition relies on strong frequency suppression in addition to high polarization purity.  
Fig.~\ref{fig:OffResonantScatt}a shows the dominant leakage channel from the \emph{bright} to the \emph{dark} F=1 manifold due to off-resonant excitation of the transition $F=2\rightarrow F'=2$ by polarization contaminated $\pi$-photons. The state $F'=2$ is separated in frequency from $F'=3$ by $\Delta_{\text{sep}}/2\pi=266\,$MHz $+\Delta_{\text{AC}}$, where the last term accounts for the AC-Stark shifts induced by the dipole trap. The ratio of excitation rates to the state $F'=2$ and hence the transfer rates to the \emph{dark} state for blue and red SDB detunings of equal magnitude $\Delta$ is then given by
\begin{equation}
\frac{R_{\text{sc}}^{\text{blue}}\left( \Delta \right)}{R_{\text{sc}}^{\text{red}}\left( \Delta \right) }\approx\left( \frac{\Delta _{\text{sep}}-\Delta}{\Delta _{\text{sep}}+\Delta }\right) ^{2},
\end{equation}
leading to a smaller leakage rate for blue detunings.\\ \indent
In order to experimentally determine transfer probabilities to the dark state, atoms initially prepared in the bright state ($F=2, m_F=-2$) are illuminated by the SDB in a trap of $k_{\text{B}} \times 3.46\,$mK depth. To count the number of atoms transferred to the state $F=1$, atoms in the state $F=2$ are removed using the \emph{push-out} technique and the remaining atoms are detected in a subsequent reference image. Fig.\ref{fig:OffResonantScatt}b shows the results of measurement taken for three different frequency detunings $\Delta_{\text{SDB}}/2\pi= -5,\,+24,\,\text{and}\,123\,$MHz, where the illumination intensities of $I/I_\text{sat}= 3.7,\ 1.2,\ \text{and}\ 3.2 $, respectively, have been chosen such that the same number of photons is detected after 17.5\,ms. The measurements confirm that the transfer to the dark state is minimized for blue detuning.

\subsection{State detection by threshold method}

\begin{figure}[t]
\centering
     \includegraphics[width=1.0\columnwidth]{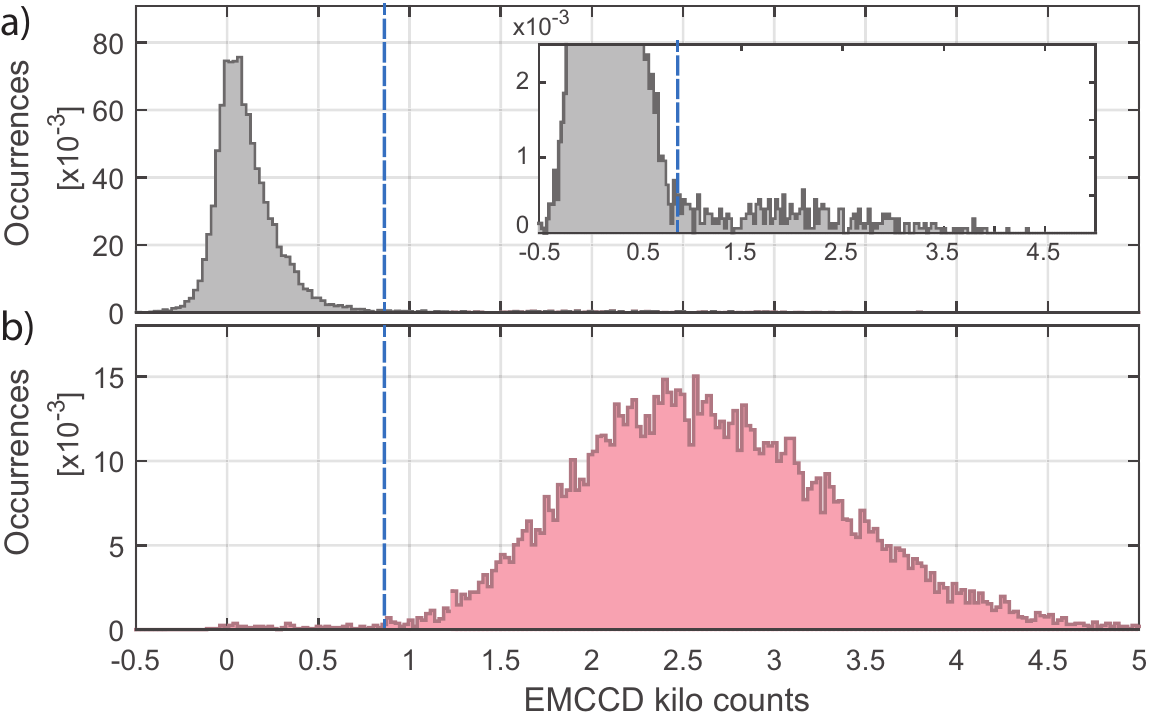}
  	\caption[Setup] {Counting statistics for single atoms during fluorescence state detection. The counts from the EMCCD camera are integrated over the region-of-interest windows as illustrated in Fig.~\ref{fig:fig1}. {\bf a)} Histogram of the number of detected counts for atoms prepared in the \emph{dark} state $F=1$. The inset highlights the low probability events, which are expected during the 10\,ms interaction time from \emph{dark} to \emph{bright} state leakage in spite of the large 2$\pi\times6.8$GHz detuning of the SDB.
{\bf b)} Histogram for atoms prepared in the \emph{bright} state $F=2, m_{F}=-2$. The EMCCD count signal for a \emph{bright} atom corresponds to an average of $\sim$~31 detected photons. The vertical blue line indicates the optimal  discriminator value for state detection by the threshold method.	 
 } {\label{fig:StateDetectionHistograms}}
\end{figure}
Combining the previous insights on heating and population leakage during photon scattering, we now employ state-dependent fluorescence imaging to perform high-fidelity non-destructive hyperfine state readout for arrays of $^{87}$Rb atoms in the optical standing wave. In our system the set of experimental parameters with $k_{\text{B}} \times 3.46\,$mK trap depth and $\Delta_\text{SDB}/2\pi= +123$\,MHz detuning and $I = 1.9 I_\text{sat}$ intensity of the SDB represents a compromise between heating and population leakage. These settings lead to the detection of an average of $\sim$~31 photons from an atom in the \emph{bright} state during 10\,ms of integration time. After the readout process, 98.4(2)\% of the atoms in the \emph{bright} and 99.1(1)\% of the atoms in the \emph{dark} state remain trapped in the same lattice site. Furthermore $\sim98\%$ of the surviving atoms remain in their initial state. Therefore, the presented method is not only non-destructive, but also preserves the internal hyperfine state.

To quantify the fidelity of the state detection method, we use the experimental sequence described in Sec.~\ref{experimentalsetup} and depicted in Fig.~\ref{fig:fig1}. The counting statistics for the \emph{bright} and \emph{dark} state shown in Fig.~\ref{fig:StateDetectionHistograms} has been compiled from the region-of-interest binning of the EMCCD counts in state detection image for 1.5$\times10^{4}$ well-resolved atoms that have been prepared in the \emph{bright} and \emph{dark} state, respectively. The optimal threshold for state discrimination $T_{\text{opt}}$ is found as the threshold value that minimizes the mean value of the state detection error~\cite{fuhrmanek2011free,gibbons2011,acton2006}
\begin{subequations}\label{eq:Errors}
\begin{align}
Err(T,|D)=\int_{T}^{\infty }P\left(c|\text{D}\right)\text{d}c,\\
Err(T,|B)=\int_{0}^{T}P\left(c|\text{B}\right)\text{d}c.
\end{align}
\end{subequations}
where $P\left(c|\text{S}\right)$ are the normalized count distributions for S=B,D. 
Applying the state detection to well-resolved atoms we find a mean detection error for the threshold method of \num{1.4(2)}\%.

\section{Conclusion}\label{Conclusions}

From the theoretical models and the experimental results here presented we conclude the following:
\begin{itemize}
\item In optical traps with weak confinement (e.g. in running-wave optical traps) the heating effects associated with the different potential of the optically excited state can be neglected compared to photon recoil.
\item In optical traps with tight confinement (e.g. in deep standing wave lattices), illumination by \emph{weak resonant} light leads to strong DFF  effects that are well described by the semi-classical absorption and emission picture. In particular for deep traps, the DFF become the dominant heating mechanism and the scattering of only a few photons can lead to the loss of the atom from the trap.
\item DFF  induced by the state detection beam are suppressed using \emph{off-resonant} illumination ($|\Delta_{1}({\bf r})|\gg\Gamma$). This is of great significance since it shows that deep optical lattices can be used to increase the number of scattered photons without atom loss. 
\item The two models implemented in the Monte Carlo simulations describe well the measured data for two experimentally relevant regimes: When resonant light creates strong DFF and when a large detuning suppresses the DFF. 
\item We have found adequate illumination settings for our particular experimental system to perform spatially-resolved, high-fidelity state-dependent imaging. The detailed understanding of the underlying DFF heating processes presented here, however, is general and can be useful for a wide range of experiments with neutral atoms in optical potentials.
\end{itemize}

\section*{Acknowledgments}
We would like to thank Klaus M{\o}lmer and Jean-Michel Raimond for the insightful discussions. This work has been supported by the Bundesministerium f\"ur Forschung und Technologie (BMFT, Verbund Q.com-Q), and by funds of the European Commission training network CCQED and the integrated project SIQS. M.Martinez-Dorantes and J.Gallego also thank the support from the Bonn-Cologne Graduate School of Physics and Astronomy.

\renewcommand\thefigure{A\arabic{figure}}   
\renewcommand\theequation{A\arabic{equation}}    
\setcounter{equation}{0} 
\setcounter{figure}{0} 
\renewcommand\thefigure{A\arabic{figure}}   
\renewcommand\theequation{A\arabic{equation}}
\setcounter{equation}{0} 
\setcounter{figure}{0} 

\section*{Appendix A. Detection efficiency}\label{sec:CollectionEfficieny}
\label{Appendix}
All the information we obtain from the atoms is provided by the detection of photons by the EMCCD camera. The \emph{collection efficiency} (CE) of a single aspheric lens (see Fig.~\ref{fig:fig1}) is obtained by the integrating the dipole emission pattern ($\sigma$ polarized in our case)~\cite{Olmschenk2009} over the solid angle subtended by the lens 
\begin{eqnarray}\label{eq:CE}
\text{CE}_{\sigma }&=&\int_{\theta =\theta _{-}}^{\theta =\theta_{+}}\int_{\phi =-\phi _{0}}^{\phi =\phi _{0}}\frac{3}{16\pi }\left( \cos
^{2}\theta +1\right) \sin \theta d\theta d\phi \\
\phi _{0}&=&\arcsin \left[ \sqrt{\frac{\text{NA}^{2}}{\sin ^{2}\theta }-\frac{1}{\tan ^{2}\theta }}\right] ,\\
\theta_{\pm}&=&\frac{\pi }{2}\pm\arcsin (\text{NA})
\end{eqnarray}
Using the numerical aperture  of our system (NA$=0.43$) Eq.~(\ref{eq:CE}) leads to a $\text{CE}_{\sigma }= 3.8\%$. The losses along the imaging path and the camera's quantum efficiency reduce the photon detection by \num{50(10)}\% leading to an overall detection efficiency of \num{1.9(5)}\%.
 
We also directly measure the detection efficiency with the following sequence. After loading a few atoms the trap is then adiabatically reduced to \SI{200}{\micro\kelvin} and switched off. The atoms are illuminated for \SI{10}{\micro\second} by  two beams: a $z-$propagating beam resonant with the cycling transition  $\left\vert 2, -2\right\rangle \rightarrow  \left\vert 2, -3 \right\rangle $ and a repumper  resonant with the transition  $F=1 \rightarrow F'=1$. The beams have an intensity of 21\,$I_{\text{sat}}$  and  14\,$I_{\text{sat}}$ respectively. We detect in average $N_\text{det ,1}=3.6$ ($N_\text{det,2}=5.2$) measured without (with) the Porro prism.

To calculate the total number of photons that the atom emits $N_\text{emit}$ in the experiment, we assume that the atom scatters photons like an ideal two-level system at a rate
\begin{equation}\label{eq:TwoLevelScattering}
R_{\text{sc}}=\left( \frac{\Gamma }{2}\right) \frac{s}{1+4\left( \Delta/\Gamma \right)^{2}+s},
\end{equation}
where $s$ is the saturation parameter and $\Delta$ the detuning of the illumination light from the atomic resonance. We assume an uncertainty of 10\% on illumination intensity and $\pm2\pi\,\times\,2.5$\,MHz on the frequency. With this considerations, the number of emitted photons is $N_\text{emit}= R_{\text{sc}} t_{\text{probe}}\approx182^{+0.8}_{-20}$. Finally, we compare the theoretical number of scatted photons and the measurement to determine detection efficiency $D_1 = N_\text{det,1}/N_\text{emit}= 1.97^{+0.11}_{-0.25}$\%  and $D_2 = N_\text{det,2}/N_\text{emit} = 2.87^{+0.07}_{-0.32}$\% using the Porro prism. The measured detection efficiency $D_1$ agrees with the expected value calculated at the beginning of this section.

\section*{Appendix B. Details of the Monte Carlo simulation}
\label{app:MCSimulation}
\subsection*{Photon scattering statistics}
The Monte Carlo simulation considers two important aspects of the photon scattering statistics in our system. First, the photon rate $R_{\text{sc}}$ is not constant due to the position dependent AC-Stark shift. Second, the multilevel structure of the atom leads to the possibility of leakage into the dark state $F=1$. To simulate these effects we proceed as follows.

\textbf{Position-dependent rates.} As shown by Zipkes et al.~\cite{zipkes2011kinetics} the random time between two scattering events can be efficienty samped, if a scattering rate has an upper bound $R_{\text{sc}}\left( \mathbf{r}\right) \leq R_{\max }$: For our model, an atom initially at a position $\mathbf{r}_0$ moves in the trapping potential during a time $\tau $ drawn from the distribution $P_{\max }=R_{\max }\exp \left( -R_{\max }t\right)$. At the position $\mathbf{r}\left( \tau \right)$ we use the re-scaled rate $g={R_{\text{sc}}\left( \mathbf{r}\left( \tau \right) \right)}/{R_{\max }}$ to decide whether the scattering event occurs by using an auxiliary random number $r \in [0,1)$ drawn from a uniform distribution and check if $r<g$. 

\textbf{Multiple scattering rates.} We consider a system where a total of $N$ independent random events can take place and each event is characterized by an exponential distribution a $\rho=R_i\exp \left( -R_i t\right)$ with rate $R_i$. We decide which event occurs by defining $\tau_{\min} =\min\{\tau _{1},\tau _{2},...,\tau_{N}\}$, where $\tau_i$ are random numbers drawn from their respective distribution.  The event i for which $\tau_{\min} =\tau_i$ is the one that takes place~\cite{van2006performance}.

\textbf{Leakage to a dark state.} In order to determine the polarization impurity of the ideally perfectly $\sigma^-$-polarized beam SDB, we tune the SDB to be resonant with the transition $F=2 \rightarrow F'=2$ and illuminate atoms initially in the $\ket{2,-2}$. From the number of atoms transferred to the state $F=1$ and their distribution over Zeeman states (which we obtain using microwave spectroscopy) the polarization impurity and its polarization components can be determined. We measure a total light impurity of $P_\text{cont}/P_\text{total}\approx 1\times10^{-3}$ and find that the polarization impurity of the light mainly consists of the $\pi$ component $(\frac{\sigma^+ }{\pi+\sigma^+}< 5\%)$. This allows us to simulate the state transfer with a simplified model: just the $\pi$ component for the polarization is considered and we assume that the events are instantaneous, i.e. the dynamics while the atom is in the ``wrong'' $m_F$ state is neglected.
 
\subsection*{Monte Carlo loop implementation for the weak resonant field}
The simulation describes a neutral atom trapped in an optical dipole trap interacting with a weak resonant field. In this case the effects of polarization contamination are neglected.

\begin{enumerate}

\item The atom is initially in the ground state $\left\vert 2, -2\right\rangle $.

\item The atom is initially at position $\mathbf{r}_{0}$ with momentum $\mathbf{p}_{\mathbf{0}}$ and a total energy $E_{0}$ drawn from a Boltzmann distribution for a given temperature $T$. 

\item \label{Item_1:CalculateRate}Calculate the maximum scattering rate $R_{\max }\geq R_\text{sc}(\Delta(r)) $ for all energy-accessible positions.

\item \label{tem_1:Draw} Draw a random time $t_1$ from the distribution $\rho=R_{max}\exp \left( -R_{max} t\right)$.

\item \label{tem_1:AdvanceSystem}  Advance the system by time $t_1$ by solving the equations of motion for $U_{g}\left( \mathbf{r}\right) $ to obtain the position and momentum, $\mathbf{r}_{1}$, $\mathbf{p}_{\mathbf{1}}$ at time $t_1$. 

\item A scattering event takes place with a probability $g={R_{sc}\left( \mathbf{r}_{1}\right) }/{R_{\max }}$. If there is a scattering event, add the photon recoil due to absorption and then continue to point~\ref{tem_1:ExciteAtom}, otherwise $\mathbf{r}_{0}=\mathbf{r}_{1}$ and $\mathbf{p}_{0}=\mathbf{p}_{1}$ and go back to point~\ref{tem_1:Draw}.

\item \label{tem_1:ExciteAtom} The atom remains in the excited for a time $t_{e}$ drawn from the distribution $\rho \left( t\right) =\Gamma \exp \left( -\Gamma t\right)$ where $\Gamma $ is the natural decay rate. The position and momentum of the atoms is updated by solving the equations of motion for $U_{e}\left( \mathbf{r}\right) $ and the photon recoil due to emission is added.

\item The simulation terminates if the energy of the atom is larger than the trap depth or if the total simulation time has reached the limit. Otherwise, the atom is again in the ground state and go back to point~\ref{Item_1:CalculateRate}.

\end{enumerate}

\subsection*{Monte Carlo loop implementation for the dressed-state potentials}
The simulation describes an atom trapped in the dressed state potential created by the interaction with a near-resonant field with an intensity $I$ that contains a small polarization contamination $I_{\pi} = I/250$. 

\begin{enumerate}
\item Initial atomic parameters: Hyperfine state is $\left\vert 2, -2\right\rangle $, Dressed state is $\left\vert -,N\right\rangle $ ( $\left\vert +,N\right\rangle $) for red- (blue-) detuning of the SDB.

\item  The atom is initially at position $\mathbf{r}_{0}$ with momentum $\mathbf{p}_{\mathbf{0}}$ and a total energy $E_{0}$ drawn from a Boltzmann distribution for a given temperature $T$. 

\item \label{itm:StartPoint} Calculate maximum rates for the current energy. These are: the decay rates $\Gamma_{\pm\pm,},~\Gamma_{\pm\mp}$ (using the subindices for the current dressed state) according to Eq.~(\ref{eq:RatesDS}) and the scattering rates for $\pi$ contamination $~R^{\pi}_{sc_{\left\vert 2,-2\right\rangle \rightarrow \left\vert 2,-1\right\rangle }},$ and $~R^{\pi}_{sc_{\left\vert 2,-2\right\rangle \rightarrow \left\vert 1,-1\right\rangle }}$ where $R^{Q}_{sc\, \left\vert\text{initial}\right\rangle \rightarrow \left\vert\text{final}\right\rangle}$ is calculated using the Kramers-Heisenberg formula for light polarizations $Q=\pi,\sigma$~\cite{takekoshi1995quasi}

\item \label{itm:DrawTimes} Draw random times from exponential  distributions for all the rates. We define $\tau$ as the minimum for the drawn times and identify the rate $R\left( \mathbf{r}\right)$ for the event that takes place and its upper bound $R_{\max}.$

\item Advance the system by a time $\tau $ by solving the equation of motion for the current dressed state potential in Eq.~(\ref{eq:DSPotentials}).

\item The scattering event takes place with a probability $g=R\left( \mathbf{r}\right) /R_{\max }.$ If the event does not take place then go back to \ref{itm:DrawTimes}.

\item Update the new hyperfine or dressed state according to the scattering event that has occurred, add the photon recoil to the atom's momentum and calculate total energy.

\item If the atomic state is $\left\vert 2,-1\right\rangle $, change the state to either $\left\vert 2,-2\right\rangle $ with probability $P_{\text{back}}$ or to $\left\vert 1, -1\right\rangle $ with probability $(1-P_{\text{back}})$, where
\begin{equation}\label{eq:Pback}
P_\text{back}=\frac{R^{\sigma}_{sc_{\left\vert 2,-1\right\rangle \rightarrow \left\vert 2,-2\right\rangle }}\left( \mathbf{r}\right) }{R^{\sigma}_{sc_{\left\vert
2,-1\right\rangle \rightarrow \left\vert 2,-2\right\rangle }}\left( \mathbf{r}\right) +R^{\sigma}_{sc_{\left\vert 2,-1\right\rangle \rightarrow \left\vert1,-1\right\rangle }}\left( \mathbf{r}\right)}.
\end{equation}

\item The simulation terminates if the energy of the atom is larger than the trap depth, or if the total simulation time has reached the limit, or if the hyperfine state is $\left\vert 1, -1\right\rangle$. If none of the previous conditions is fulfilled, then go to point \ref{itm:StartPoint}.

\end{enumerate}

\bibliography{PRAPotential}
\bibliographystyle{apsrev4-1}

\end{document}